\documentclass{ws-mpla-arXiv}

\begin{document}

\markboth{Christos Leonidopoulos}
{First Results from the CMS Experiment}

\newcommand {\etal}{\mbox{et al.}\,} 
\newcommand {\ie}{\mbox{i.e.}\,}     
\newcommand {\eg}{\mbox{e.g.}\,}     
\newcommand {\etc}{\mbox{etc.}\,}     
\newcommand {\vs}{\mbox{\sl vs.}\,}      

\newcommand{\inb}{nb$^{-1}$\,}
\newcommand{\ipb}{pb$^{-1}$\,}
\newcommand{\ifb}{fb$^{-1}$\,}
\newcommand{\imub}{$\mu$b$^{-1}$\,}

\newcommand{\MET}{\ensuremath{E_{\mathrm{T}}^{\mathrm{miss}}}\,}

\catchline{}{}{}{}{}

\title{First results from the 2010 $pp$ Run and \\ performance of the CMS Experiment\footnote{This article is based on HEP seminars given at Harvard University (January 2010) and Stanford University (February 2011) describing the performance of the CMS detector and reviewing the first physics results from the 2010 $pp$ Run at the LHC.} ~\footnote{Preprint of an article submitted for consideration in Modern Physics Letters A \copyright\ 2011 [copyright World Scientific Publishing Company, {\tt http://www.worldscientific.com/}].}}

\author{\footnotesize CHRISTOS LEONIDOPOULOS}

\address{Fermi National Accelerator Laboratory \\
P.O. Box 500, Wilson Road \\
Batavia, IL 60510, USA\\
Christos.Leonidopoulos@cern.ch}

\maketitle

\pub{May 2011}{}
\begin{abstract}
In 2010, the Compact Muon Solenoid (CMS) experiment at LHC recorded over 45 \ipb of $pp$ collision data at $\sqrt{s}$=7 TeV. The large collected datasets are of very high quality and have been used to commission and calibrate the CMS detector, with the achieved performance close to the TDR specifications. CMS has reestablished all the major Standard Model processes in the 2010 Run and is entering new territory in searches for New Physics, with sensitivity already exceeding that at LEP and TeVatron.
\end{abstract}


\section{The CMS detector}
CMS \cite{bib:cms} is one of the two general purpose particle detector experiments located at CERN's Large Hadron Collider (LHC). The central features of the CMS apparatus are the superconducting solenoid with an internal diameter of 6 m producing a uniform magnetic field of 3.8 TeV, and the muon tracking system which makes use of the steel return yoke of the magnet. Immersed in the magnetic field are
\begin{itemize}
\item the inner tracker, consisting of three layers of silicon pixel detectors (with 66 million channels) and ten layers of silicon microstrips (with 9.6 million channels), with a total active area of 210 m$^2$. The tracker allows the reconstruction of charged particles up to pseudorapidity ($|\eta|$) values of 2.5 and provides an impact parameter resolution of 100 $\mu$m and a vertex position resolution of 15 $\mu$m.
\item the electromagnetic calorimeter, made of approximately 76,000 lead-tungstate scintillating
crystals, providing an energy resolution of about 0.5\% for high energy electromagnetic showers.
\item a hermetic hadron calorimeter covering the $|\eta| < 5.2$ region, made of approximately 7,000 plastic scintillator tiles and brass absorber plates. 
\end{itemize}
Outside the solenoid is the muon spectrometer. It employs twelve layers of Drift Tubes in the barrel and six layers of Cathode Strip Chambers in the endcap, combined with Resistive Plate Chambers spanning the full detector for redundant muon coverage. Muon tracks with measurements in the silicon tracker and the muon subdetectors have excellent transverse momentum resolution, which ranges from 1\% for low-$p_T$ tracks up to $\sim$10\% for $p_T$ values around 1 TeV/$c$.
\section{The CMS Trigger}
The CMS online selection system carries out the usual gradual reduction of the background rate in two steps. The first step is the Level-1 system (L1), made of custom hardware and low-level firmware,\cite{bib:l1} which brings the LHC 40 MHz clock rate down to 100 kHz. For the higher-level filtering, instead of following the traditional trigger design with separate steps for the Level-2 and Level-3 components, CMS has followed a novel approach: it has merged these two steps into a single entity called the High-Level Trigger (HLT).\cite{bib:hlt1,bib:hlt2} The HLT runs advanced selection algorithms, which can be of offline quality at the full detector granularity. It is implemented in a “farm” comprising $\sim$5000 commercial CPUs. It provides the flexibility of a continuous software environment accessing the full L1 accept rate (\ie 100 kHz), at the expense of a large data-throughput and a significantly increased complexity of the software. The HLT has been extremely robust and reliable during the 2010 Run, recording more than one billion physics events with minimum downtime.

 The versatility of the HLT has allowed CMS to record large datasets with unusual topologies that would be very hard, if not impossible, to collect with a traditional HEP trigger. One such example is the implementation of detector-calibration triggers that run at the HLT and have, therefore, access to a very large L1 rate. The peculiarity of these triggers is that they have been designed to record only a small fraction of the detector event information. So, they can be tuned to accept a very large number of events for a relatively low data throughput. In the particular case of the $\pi^0$ calibration trigger, the stored information is limited to the kinematic information of diphoton candidates. This has allowed CMS to identify and reconstruct $\pi^0$s in real time at a rate of about 100 times higher than what would be possible with a traditional HEP trigger, providing very large statistics for the commissioning of the calorimeter. In effect, the detector calibration process starts online and this is a feature that is unique to CMS and a first for hadron collider triggers.

Other examples that exploit the novel HLT design are the customized triggers developed to capture stopped gluinos and high-multiplicity events for the study of the so-called ridge-effect, and are briefly discussed later in this article.
\section{First LHC Runs and data collected}
In the pilot $pp$ collision Runs at the end of 2009, CMS recorded approximately 10 \imub of data at $\sqrt{s}$ = 900 GeV and 0.5 \imub of data at $\sqrt{s}$ = 2.36 TeV. The collected datasets consisted of about 350,000 and 20,000 minimum bias events, respectively, and were used to commission the detector and tune the physics object reconstruction. The $\sqrt{s}$ = 7 TeV $pp$ run started in April 2011 and lasted for seven months, with LHC reaching a record instantaneous luminosity of 2 $\times 10^{32}$ cm$^{-2}$\,s$^{-1}$. CMS collected more than one billion physics events, corresponding to about 45 \ipb of integrated luminosity. Starting in November 2011, LHC delivered lead-lead collisions at $\sqrt{s_{NN}}$ = 2.76 TeV for about one month, and CMS collected approximately 10 \imub of heavy-ion collision events.

In the following sections we are reviewing some highlights from the first physics results of the 2010 $pp$ Run at $\sqrt{s}$ = 7 TeV.
\section{The Particle-Flow reconstruction}
\label{sec:pf}
The Particle-Flow (PF) is a full-event reconstruction technique, originally developed at LEP \cite{bib:pf_lep} and widely used at CMS.\cite{bib:pf} It aims at reconstructing and identifying all stable particles in the event (\ie electrons, muons, photons, charged and neutral hadrons) by combining information across subdetectors for the optimal determination of their kinematic properties and minimal fake rate. The minimum requirements for a well-performing PF reconstruction at the challenging topologies of LHC collision events are
\begin{itemize}
\item an excellent tracker system for reliable $p_T$ measurements
\item a fine-granularity electromagnetic calorimeter for associating tracks to energy clusters with a very small fake rate, and
\item a strong magnetic field for disentangling the very large number of charged tracks when associating them with particle traces in other subdetectors.
\end{itemize}
The CMS detector has all these features and is, therefore, very well suited for PF reconstruction.

\begin{figure}[htb]
\centerline{\psfig{file=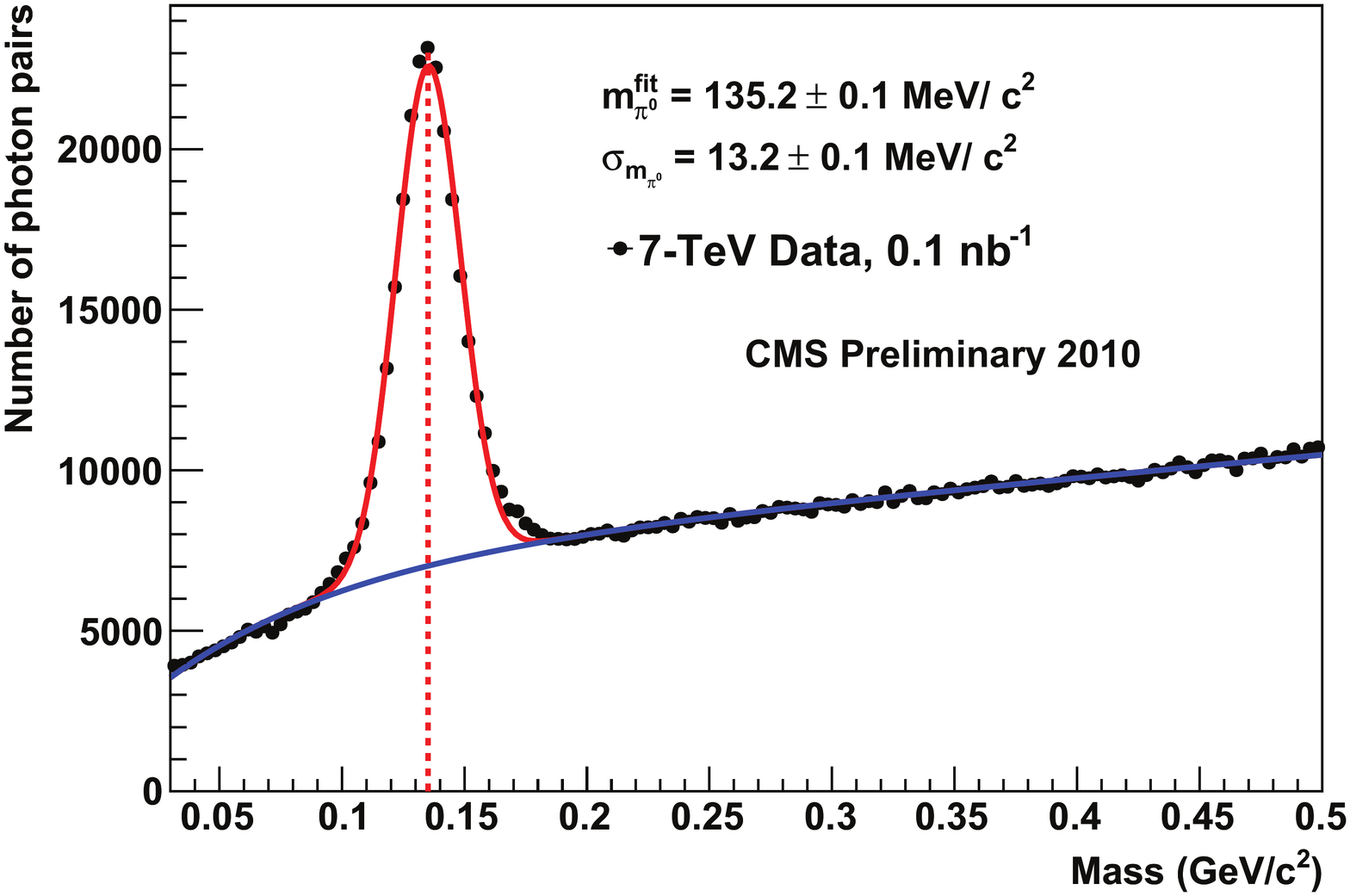, angle=90, width=4.0in}}
\vspace*{8pt}
\caption{The diphoton mass distribution reconstructed with the PF algorithm with the first 0.1 \inb of collision data. Simulation-based corrections have been applied in the reconstruction of calorimetric clusters and photon candidates. \protect\label{fig:pfPi0}}
\end{figure}

\begin{figure}[htb]
\centerline{\psfig{file=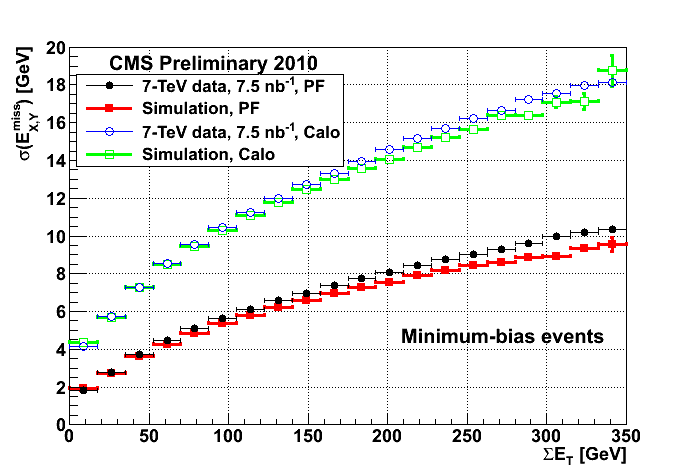,width=4.0in}}
\vspace*{8pt}
\caption{Missing transverse energy resolution for calorimetric-based (open symbols) and PF (solid symbols) reconstruction in simulation (squares) and data (circles) as a function of the transverse-energy sum in the event in 7.5 \inb of minimum bias events.\protect\label{fig:pfMET}}
\end{figure}

As examples of the excellent PF performance, we present two plots from the early analysis of the 2010 data. Fig.~\ref{fig:pfPi0} shows the diphoton mass distribution in the first 0.1 \inb of data, as reconstructed with the PF algorithm, with a clearly visible $\pi^0$ mass peak. A fit on the peak reproduces the world average value for the $\pi^0$ mass, despite the fact that only simulation-based calibration and corrections were applied in the calorimetric cluster reconstruction at the time. Fig.~\ref{fig:pfMET} shows the comparison of the missing transverse energy (\MET) resolution for calorimetric-based and PF reconstruction with 7.5 \inb of minimum bias events. The PF reconstruction improves the calorimetric \MET measurement by a factor of two. The same plot shows a very good agreement between the simulated and actual performance of the \MET measurement, demonstrating a good level of detector calibration and understanding.

\section{Rediscovery of the Standard Model (SM)}

With the 2010 data CMS has completed the commissioning of its detector and physics object reconstruction, and carried out preliminary measurements of key SM processes: examples include the published results on the transverse momentum and pseudorapidity distributions of charged hadrons,\cite{bib:chd_had} and the cross-section determination of the inclusive jet,\cite{bib:inc_jet}  inclusive photon,\cite{bib:inc_pho} heavy-mesons,\cite{bib:chd_B,bib:neu_B,bib:Bs} onia\cite{bib:onia1,bib:onia2,bib:onia3}, electroweak\cite{bib:ewk} and top\cite{bib:top} production. These measurements constitute an important step for CMS in re-establishing the known collider physics before exploring the new energy regime in searches for physics beyond the SM.

In the following sections we are discussing the cross-section measurements of the $W$, $Z$ and top production, as representative results of the large number of SM analyses that CMS carried out with the 2010 data.

\subsection{$W/Z$ production}
\label{sec:WZ}
The importance of the $W$ and $Z$ boson production measurements at the LHC (that correspond to clean theoretical calculations and can therefore be used as standard ``reference candles'') has been discussed extensively in the literature.\cite{bib:candle} CMS's startup strategy has been to focus on the leptonic channels (electrons and muons in particular) that are relatively easy to reconstruct and can provide quick measurements. 

\begin{figure}[htb]
\begin{center}$
\begin{array}{cc}
\includegraphics[width=2.5in]{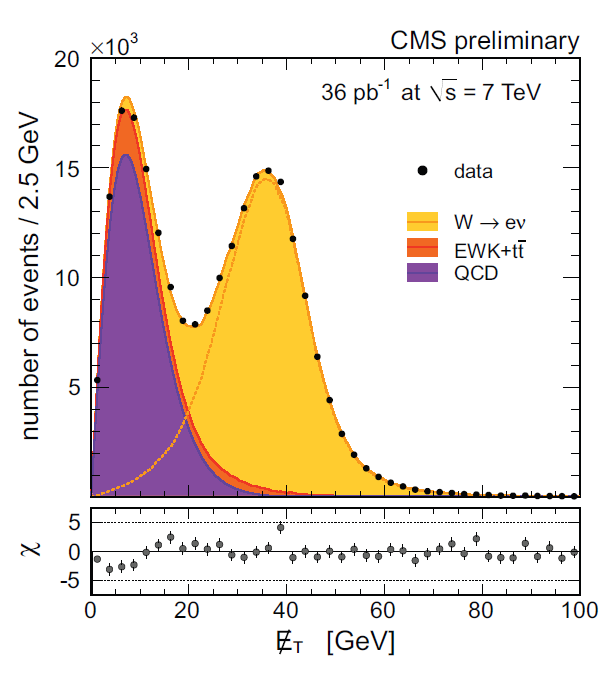} &
\includegraphics[width=2.5in]{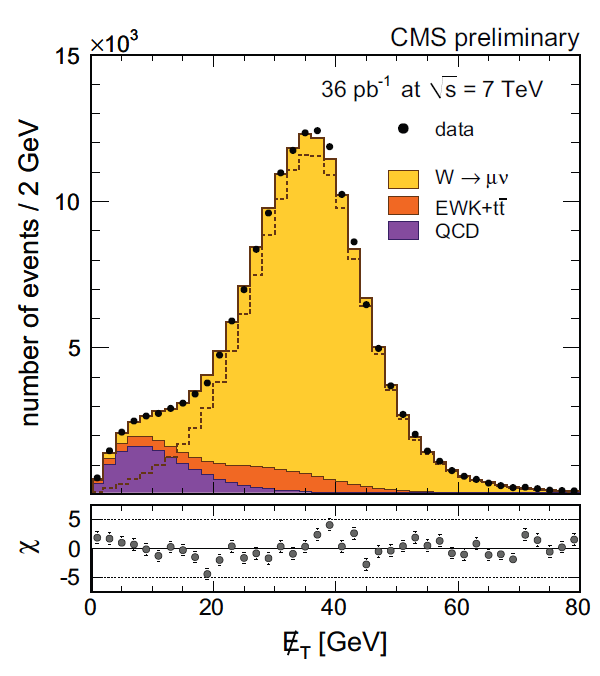}
\end{array}$
\vspace*{8pt}
\end{center}
\caption{The \MET distributions for the $W \rightarrow e \nu$ (left) and $W \rightarrow \mu \nu$ (right) candidates for 36 \ipb.The points represent the data. Superimposed are the results of the maximum likelihood fits for
signal plus backgrounds, in yellow; all backgrounds, in orange; QCD backgrounds, in violet.\protect\label{fig:W}}
\end{figure}

\begin{figure}[htb]
\begin{center}$
\begin{array}{cc}
\includegraphics[width=2.5in]{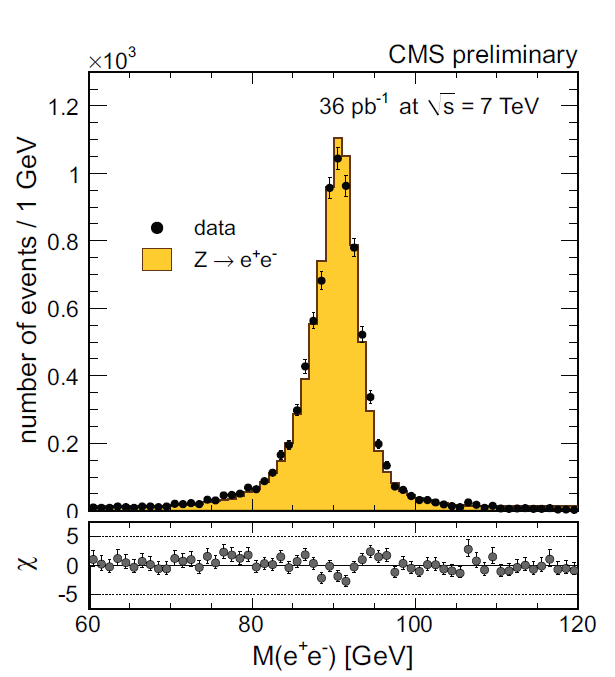} &
\includegraphics[width=2.5in]{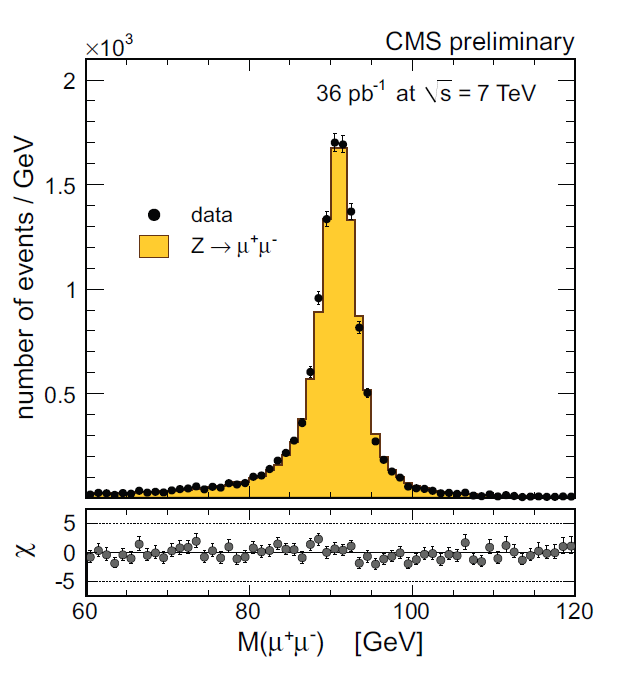}
\end{array}$
\vspace*{8pt}
\end{center}
\caption{The invariant mass distributions for the $Z \rightarrow e^+ e^-$ (left) and $W \rightarrow \mu^+ \mu^-$ (right) candidates for data (dots) and Monte Carlo (histogram) signal and background events for an integrated luminosity of 36 \ipb.\protect\label{fig:Z}}
\end{figure}

Good quality electrons and muons are selected. An isolation requirement is further applied, in order to reduce background contributions from leptons contained in jets. For the $W$ cross section measurement the \MET is calculated with the PF technique (see Sec.~\ref{sec:pf}) for events with one good lepton identified. The number of signal events is estimated by exploiting its distinct shape in the \MET distribution (Fig.~\ref{fig:W}). For the $Z$ cross section measurement events with two good leptons ($e^+ e^-$, $\mu^+ \mu^-$) are selected, and their invariant mass is reconstructed (Fig.~\ref{fig:Z}). The background here is negligible and is estimated from simulation. 

We measure the following cross sections for the $W$, $Z$ production, and their ratio:
{\small
\begin{align*}
\sigma (pp \rightarrow WX) \times BF (W \rightarrow \ell \nu) &= 10.31 \pm 0.02~({\rm stat}) 
\pm 0.09~({\rm syst}) \pm 0.10~({\rm th})~{\rm nb} \\
\sigma (pp \rightarrow ZX) \times BF (Z \rightarrow \ell^+ \ell^-) &= 0.975 \pm 0.007~({\rm stat}) 
\pm 0.007~({\rm syst}) \pm 0.018~({\rm th})~{\rm nb} \\
{{\sigma (pp \rightarrow WX) \times BF (W \rightarrow \ell \nu)}\over {\sigma (pp \rightarrow ZX) \times BF (Z \rightarrow \ell^+ \ell^-)}} &=10.54 \pm 0.07~({\rm stat}) \pm 0.08~({\rm syst}) \pm 0.16~({\rm th})
\end{align*}
}
The first two measurements have an additional integrated luminosity uncertainty of 4\%, which cancels out when taking their ratio. All measurements are in good agreement with the NNLO predictions.\cite{bib:ewk}

\subsection{Top production}
The top is probably the most interesting of the SM quarks, with a mass much heavier than all other known quarks or leptons and observed only at the TeVatron until very recently. Its significantly enhanced production rate at the LHC is expected to facilitate in-depth studies of its properties. It is also a prominent decay product in many New Physics signatures, so its efficient reconstruction and determination of its production rate is considered very important.

\begin{figure}[htb]
\centerline{\psfig{file=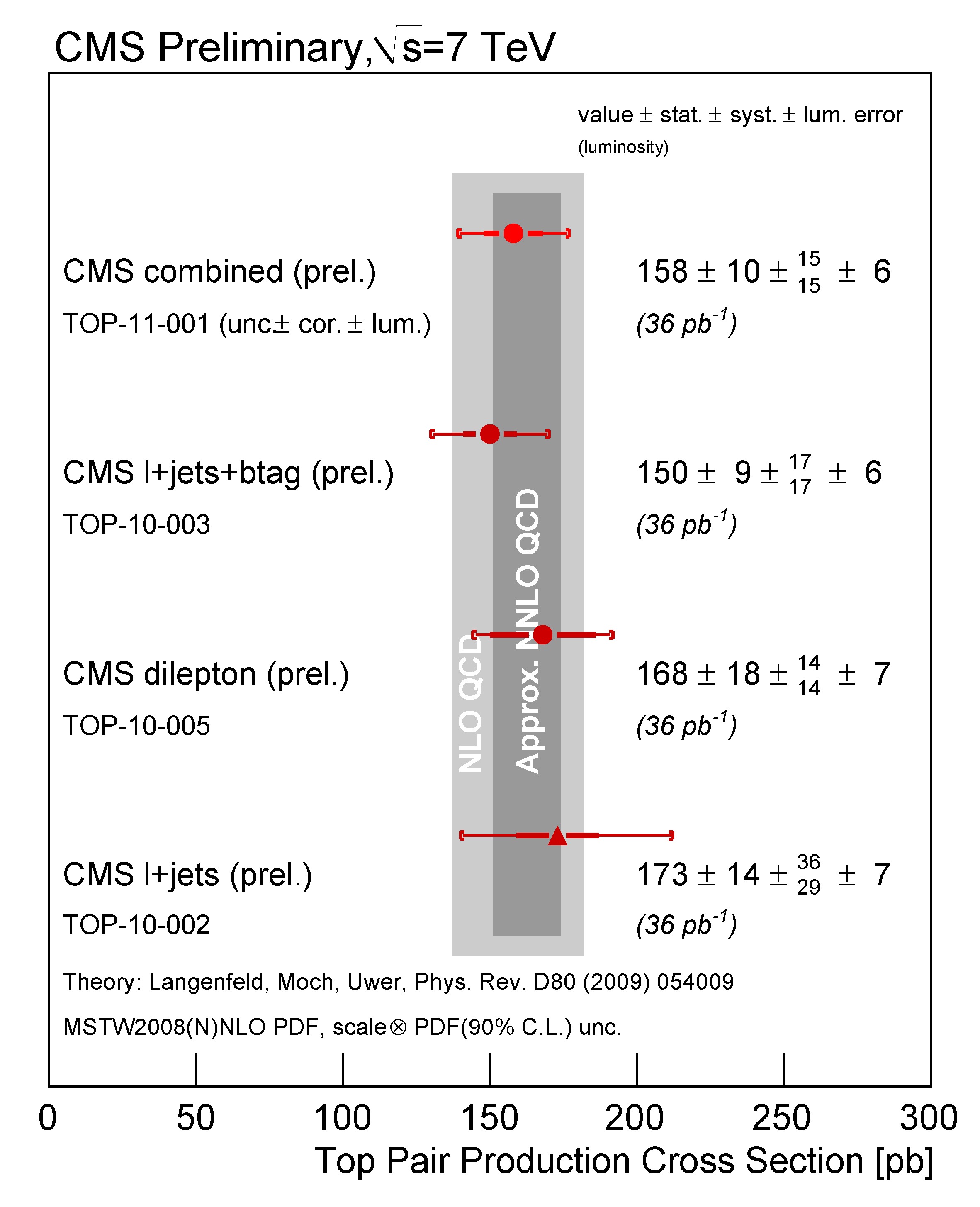, width=3.0in}}
\vspace*{8pt}
\caption{Summary of CMS inclusive $t\bar{t}$ cross section measurements at $\sqrt{s}$ = 7 TeV with 36 \ipb of data. \protect\label{fig:top}}
\end{figure}

CMS carried out in 2010 several analyses for the evaluation of the $t\bar{t}$ cross-section in both dilepton and lepton + jets  topologies (with and without $b$-tagging), in the electron and muon flavors. A summary of these measurements can be found in Fig.~\ref{fig:top}. An overall good agreement is observed with theory predictions at NLO and approximate NNLO.\cite{bib:top}
\section{Exotic searches}
By the term ``exotic searches'', we typically mean searches for new, yet-to-be-discovered, heavy particles decaying into known matter. For this type of searches, the large increase in the collision energy in the transition from the TeVatron (1.96 TeV) to the LHC (currently 7 TeV, eventually 14 TeV) is the single most important factor that determines the discovery potential of the LHC experiments.

In the next sessions we are reviewing the results of searches for leptoquarks, new vector bosons and long-lived particles that have been carried out with the 2010 data. 

\subsection{Leptoquarks}
Leptoquarks are hypothetical particles predicted in Grand Unifying Theory models that carry both lepton and baryon numbers. They provide a natural explanation as to why there are as many leptons as quarks. They could also offer a possible answer to the baryogenesis question, if combined with baryon number violation and $CP$ violation in the leptoquark decay. Leptoquarks are typically assumed to be produced in pairs which couple to single lepton-quark generations, in order to avoid flavor-changing currents. They decay to leptons and quarks, which are detected experimentally as jets that materialize inside the detector. 

 In 2010, CMS carried out searches in the first (electron) and second (muon) leptoquark generations, and observed no deviation from the expected SM background. The variable employed to set an exclusion limit is the scalar sum ($S_T$) of the transverse momenta of all final products, namely the two leptons and the two jets. This variable is not affected by combinatorics and only moderately affected by initial or final state radiation. With the presently available data, it offers an increased sensitivity compared to the invariant mass of the two leptoquark candidates. The main backgrounds in this analysis are Drell-Yan plus jets, and $t\bar{t}$ to a lesser extent.  

\begin{figure}[htb]
\begin{center}$
\begin{array}{cc}
\includegraphics[width=2.5in]{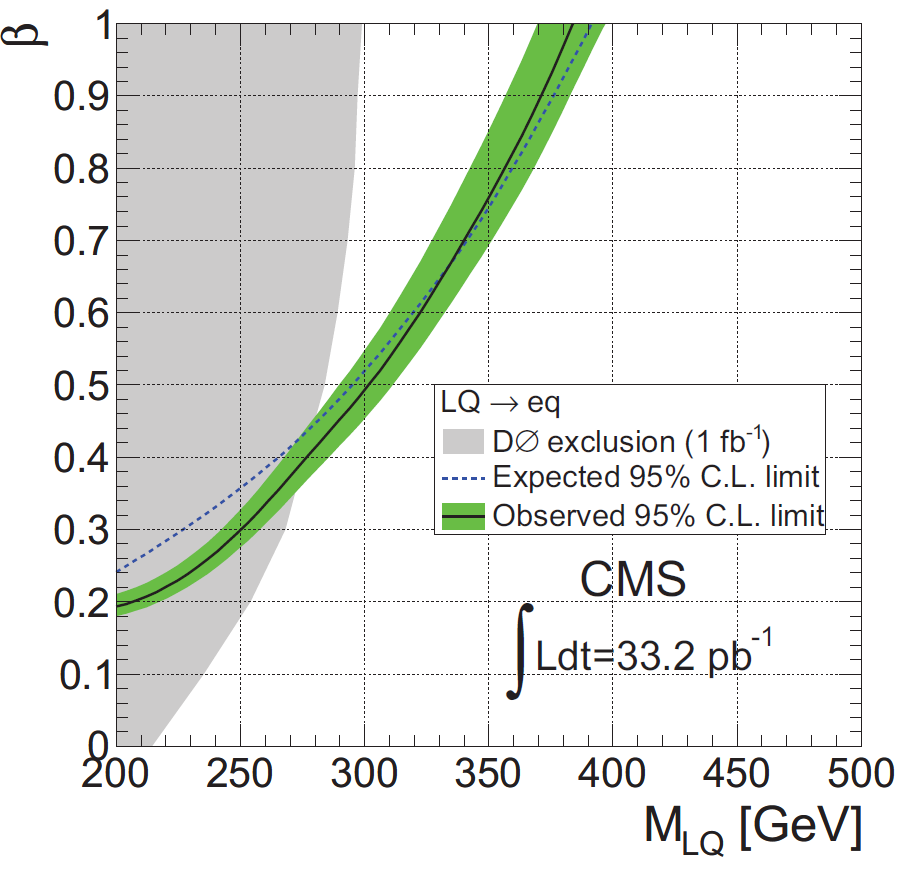} &
\includegraphics[width=2.5in]{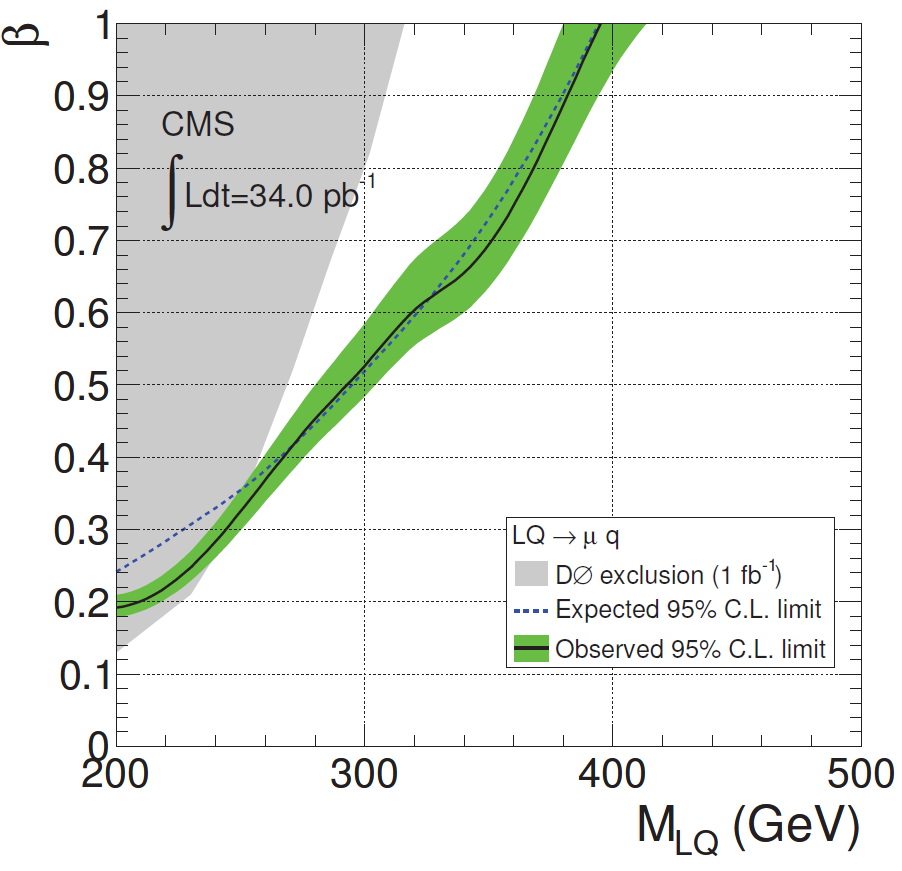}
\end{array}$
\vspace*{8pt}
\end{center}
\caption{Minimum $\beta \equiv BR(LQ \rightarrow \ell q)$ for leptoquark (LQ) hypothesis exclusion \vs mass for first ($eejj$, left) and second ($\mu\mu jj$, right) generation leptoquark searches. Previous exclusion regions established by the D0 experiment are also shown, which combined results from the $\ell \ell jj$,   $\ell \nu jj$\ and  $\nu \nu jj$ channels. \protect\label{fig:lq}}
\end{figure}

Fig.~\ref{fig:lq} shows the minimum $\beta \equiv BR(LQ \rightarrow \ell q)$ for excluding at 95\% C.L. the leptoquark hypothesis as a function of its mass for first (left) and second (right) generation searches. The corresponding experimental signatures for the two channels are $eejj$ and $\mu\mu jj$.  The limits set by these analyses are the strictest in the world to date.\cite{bib:lq1,bib:lq2} 

\subsection{Heavy vector bosons: $W^\prime$, $Z^\prime$}
Several beyond-the-SM theories predict new heavy vector bosons, as an extension to the SM gauge group. In some of these models the new particles are treated as carbon copies of the SM $W$ and $Z$ bosons,\cite{bib:altarelli} even though there are also more exotic models that introduce right-handed couplings.

Experimentally, the signatures range from the simplest ones that are encountered in the SM vector boson decays (\ie a lepton and \MET for the $W^\prime$, a dilepton for the $Z^\prime$ and light quarks for both bosons) to new channels that become kinematically available because of the assumed large mass of the new boson ($t\bar{b}$  for the $W^\prime$ and $t\bar{t}$ for the $Z^\prime$).  The searches in 2010 concentrated on the leptonic channels in the electron and muon flavors.

The analyses for the $W^\prime$ and $Z^\prime$ searches are very similar to the ones for the $W$ and $Z$ cross-section measurements (see Sec.~\ref{sec:WZ}) with small modifications. The muon reconstruction and the electron identification have been tuned to deal with very energetic leptons. In the $W^\prime$ case, a requirement that the directions of the lepton and the \MET are back to back in the transverse plane reduces the $W$+jets contamination, as a heavy $W^\prime$ would leave very little phase space for jet radiation. The remaining background consists of predominantly off-peak (heavy) $W$s and is, therefore, irreducible. In the $Z^\prime$ case, in order to reduce the cosmic background that could appear as muon pairs in the top and the bottom halves of the detector, a requirement that the two muons are not back to back in the transverse plane is applied. Here the dominant, and also irreducible, background is the Drell-Yan process. 

\begin{figure}[htb]
\begin{center}$
\begin{array}{cc}
\includegraphics[width=2.5in]{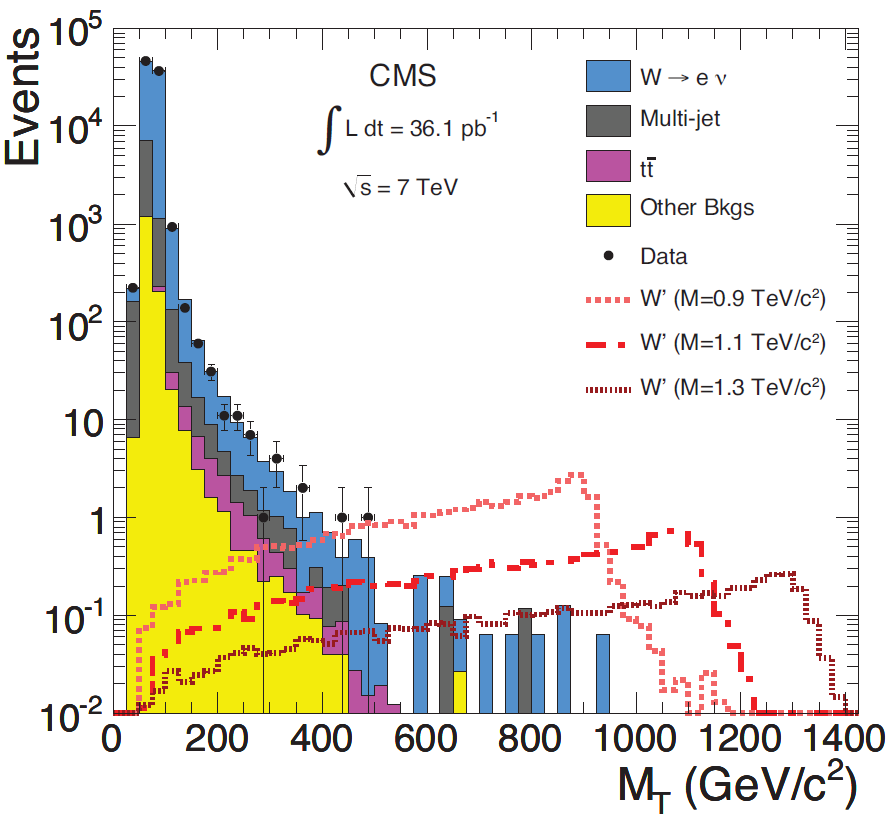} &
\includegraphics[width=2.5in]{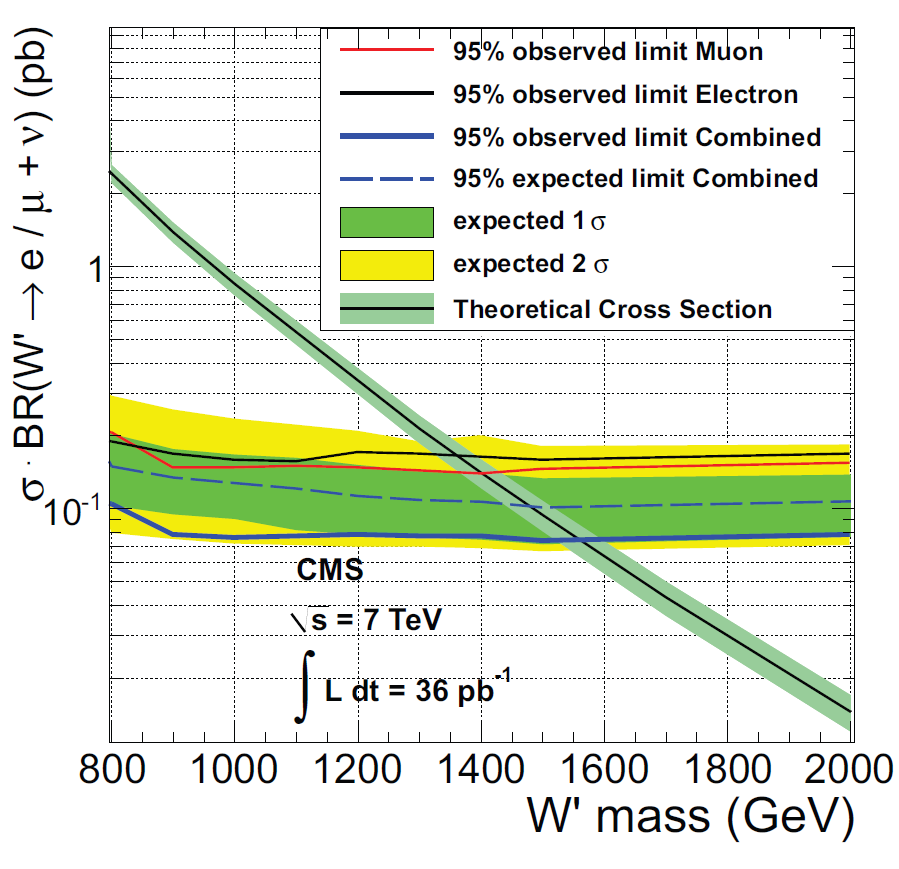}
\end{array}$
\vspace*{8pt}
\end{center}
\caption{Left: Electron and \MET transverse mass distribution for data and simulated background for an integrated luminosity of 36 \ipb. The characteristic Jacobian peaks correspond to simulated $W^\prime$ particles of various mases. Right: 95\% C.L. exclusion limits for the electron and the muon channels and their combination. \protect\label{fig:wprime}}
\end{figure}

The leftmost plot in Fig.~\ref{fig:wprime} shows the electron and \MET transverse mass distribution for data and simulated background, and what a potential $W^\prime$ signal with different mass hypotheses would look like. The rightmost plot in the same figure shows the exclusion limits for the electron and muon channels and their combination. $W^\prime$ particles with masses up to 1.58 TeV/$c^2$ in the sequential SM model are excluded at 95\% C.L., setting the most stringent limit in the world.\cite{bib:wprime_e,bib:wprime_mu}  

\begin{figure}[htb]
\centerline{\psfig{file=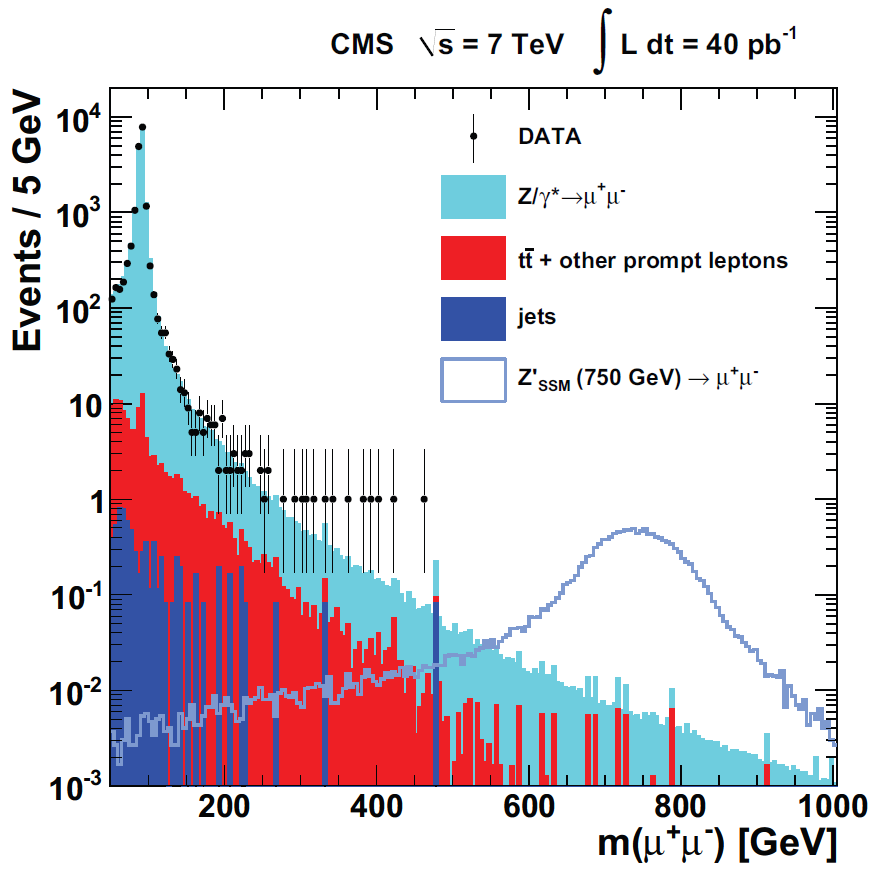, width=3.0in}}
\vspace*{8pt}
\caption{Invariant mass spectrum of $\mu^+ \mu^-$ events for data and simulated background for an integrated luminosity of 40 \ipb, along with the expected signal for a $Z^\prime_{\rm SSM}$  with a mass of 750 GeV$/c^2$. \protect\label{fig:zprime_mumu}}
\end{figure}

\begin{figure}[htb]
\centerline{\psfig{file=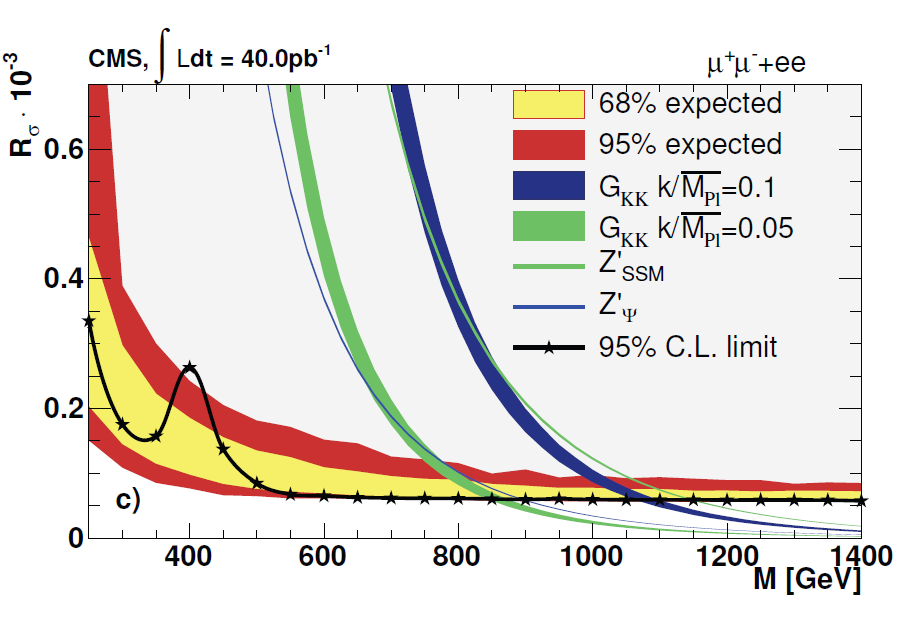, width=4.0in}}
\vspace*{8pt}
\caption{Upper limits as a function of resonance mass on the ratio of the $Z^\prime / Z$ of cross section times branching fraction into lepton pairs for various $Z^\prime_{\rm SSM}$, $Z^\prime_{\rm \psi}$ and Kaluza-Klein models.\protect\label{fig:zprime_limits}}
\end{figure}

Fig.~\ref{fig:zprime_mumu} shows the invariant dimuon mass spectrum for data and simulated background and a simulated $Z^\prime$ signal with a mass of 750 GeV$/c^2$ in the sequential SM framework. Fig.~\ref{fig:zprime_limits} shows the exclusion limits as a function of the $Z^\prime$  mass for various $Z^\prime_{\rm SSM}$, $Z^\prime_{\rm \psi}$ and Kaluza-Klein models. These limits are comparable to, or already exceed those published from previous search results at the TeVatron.\cite{bib:zprime}
\subsection{Long-lived particles}
The term long-lived particles refers to hypothetical particles with unusually large lifetimes. Examples are gluinos in ``split'' SUSY that are much lighter than squarks, light stops with very limited decay modes, staus as NLSPs that decay only via gravitational couplings, hidden valley models, \etc The detection of these particles presents experimental challenges as they tend to decay out-of-time with respect to the main interaction, or they leave tracks that do not point to the interaction point. 

There are two broad analysis strategies. The traditional one is the search for massive charged particles that leave highly ionizing tracks in the silicon tracker and the muon subdetectors. The experimental signature is similar to that of a ``heavy muon'' with large associated $dE/dx$ values. The second approach (and what is discussed in this article) is the search for strongly interacting, but slower, particles that decay inside the detector one or more bunch crosses later than the main interaction. Most of the sensitivity comes from particles that are so slow that they come to rest. This occurs in the densest regions of the detector: a large fraction of them ($\sim$55\%) would stop in the calorimeter. The experimental signature is a jet in coincidence with a bunch empty of protons. The complicated beam structure at the LHC with various beam-gap sequences gives the opportunity for a wide range of lifetimes coverage. 

\begin{figure}[htb]
\centerline{\psfig{file=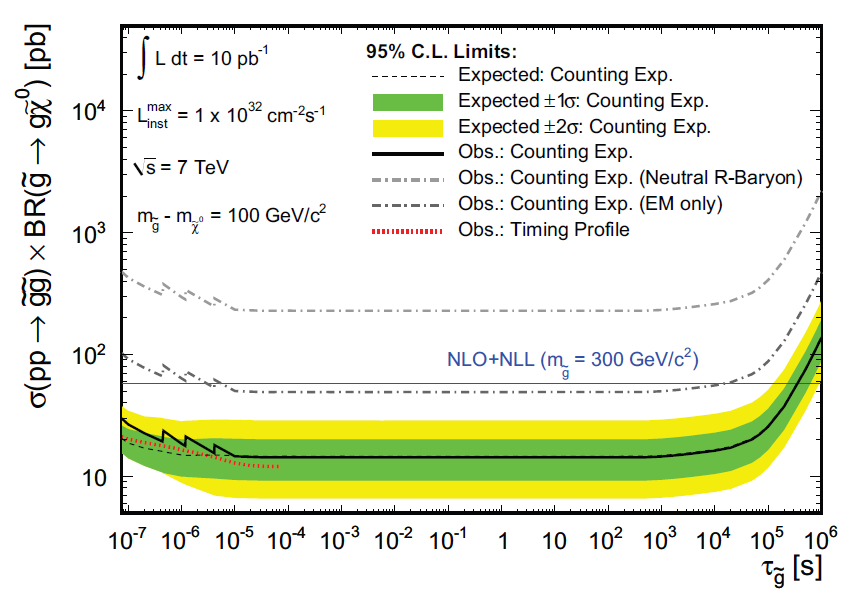, width=3.5in}}
\vspace*{8pt}
\caption{Expected and observed 95 \% C.L. limits on gluino pair production cross section times branching fraction using the “cloud model” of $R$-hadron interactions as a function of gluino lifetime. Observed 95 \% C.L. limits on the gluino cross section for alternative R-hadron interaction models are also presented. The NLO+NLL calculation is for a gluino mass of 300 GeV/$c^2$ . \protect\label{fig:gluino}}
\end{figure}

To capture this unusual signal, a customized trigger was designed requiring a jet in anti-coincidence with a filled bunch. This allowed CMS to explore the vast ``dead'' regions of empty bunches that made up the majority of the beam structure in 2010, and significantly extend the discovery potential for stopped gluinos. No excess of events above the expected background is observed. In order to place an upper limit on the process cross-section, several assumptions on the hadronization of the gluino and the fragmentation of the hadron are made, which are generally model-dependent and introduce large systematic uncertainties in the analysis. The results are summarized in Fig.~\ref{fig:gluino}. The baseline is the so-called cloud model and only scenarios in which the gluino mass is at least 100 GeV$/c^2$ larger than the neutralino mass are considered. As an example, by setting the gluino mass to 300 GeV$/c^2$, CMS can exclude lifetimes from 75 ns to 300,000 sec, \ie in a range spanning over 13 orders of magnitude. These are the most stringent limits in the world in the production of stopped gluinos to date.\cite{bib:stopped_gluino}

\section{Other results: The ``Ridge'' Effect}
Early in the 2010 $pp$ Run, CMS deployed another dedicated trigger for collecting high-multiplicity events. One of the motivations was to study two-particle correlation effects in high-multiplicity environments for comparison with results from heavy-ion collisions at RHIC.\cite{bib:rhic} The requirement at L1 was a simple energy-sum (initially at 65 GeV, eventually raised to 80 GeV). At HLT, an algorithm reconstructing and counting silicon pixel tracks for a large fraction of the L1 bandwidth was engineered, bypassing completely intermediate rejection steps based on calorimetric and muon information. Even though this filtering is very CPU-intensive, it can yield high-multiplicity datasets that can be up to 1000 times larger than with a traditional trigger.

\begin{figure}[htb]
\begin{center}$
\begin{array}{cc}
\includegraphics[width=2.5in]{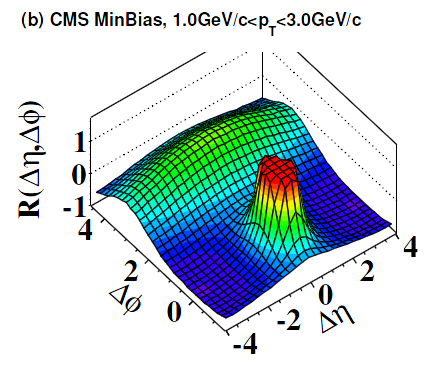} &
\includegraphics[width=2.5in]{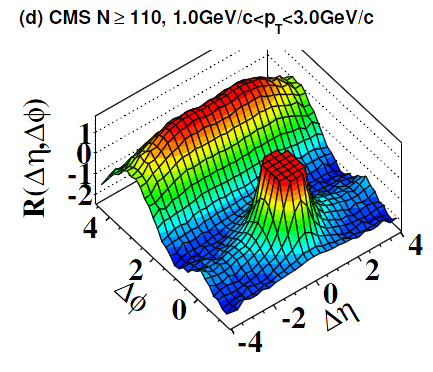}
\end{array}$
\vspace*{8pt}
\end{center}
\caption{2-D two-particle correlation functions for 7 TeV $pp$ minimum bias events with $1 < p_T< 3$ GeV/$c$ (left)  and high multiplicity ($N_{trk}> 110$) events with $1 < p_T< 3$ GeV/$c$ (right). The sharp near-side peak from jet correlations is cut off in order to better illustrate the structure outside that region. \protect\label{fig:ridge}}
\end{figure}

The analysis considers the $\Delta \phi$, $\Delta \eta$ angular distributions for all particle pairs. Fig.~\ref{fig:ridge} on the left shows the usual correlations of particles within jets around ($\Delta \phi$, $\Delta \eta$) $\sim$ (0,0) and the correlations in particles in dijets that lie in opposite directions in the transverse plane in minimum bias events. For the high-multiplicity datasets (Fig.~\ref{fig:ridge}, right) one observes correlations in particle pairs at large $\Delta \eta$ values, corresponding to particles which are receding from each other while traveling along the same $\phi$ angle. This was the first such observation at LHC with $pp$ collisions\cite{bib:ridge} and it resembles a similar feature reported by RHIC in heavy-ion Runs with gold collisions.\cite{bib:rhic} This analysis serves as a testament to the versatility of the CMS trigger that allowed the observation of a very interesting and unexpected result by recording unusual topologies in large statistics.
\section{Higgs discovery prospects}
As an epilogue, we present the expectations for the outcome of the Higgs searches in the ongoing 2011-12 $pp$ Run. These projections have been made by using the actual detector performance and an expanded list of decay channels:\cite{bib:higgs} $H \rightarrow WW \rightarrow 2 \ell \,2 \nu$ (+0/1 jets), $H \rightarrow ZZ \rightarrow 4 \ell$,  $H \rightarrow ZZ \rightarrow 2 \ell \, 2\nu$,  $H \rightarrow ZZ \rightarrow 2 \ell \,2b$, $H \rightarrow \gamma \gamma$, $H \rightarrow \tau \tau$, $VH \rightarrow V(bb)$, $ZH \rightarrow Z(WW) \rightarrow (\ell \ell) (\ell \nu jj)$, $WH \rightarrow W(WW) \rightarrow (\ell \nu) (\ell \nu jj)$. The results are summarized in Figs.~\ref{fig:higgs_limits} and \ref{fig:higgs_sign}. If there is no Higgs, CMS should be able to exclude it at 95\% C.L. for masses between 130 and 450 GeV/$c^2$ with 1 \ifb of data at $\sqrt{s}$ = 7 TeV.  If there is a SM Higgs, CMS should be able to make a 5$\sigma$ discovery (4$\sigma$ observation) for masses between 140 (130) and 240 (500) GeV/$c^2$ with 5 \ifb of data at $\sqrt{s}$ = 7 TeV. Conservative estimates indicate that CMS will have recorded this integrated luminosity before the end of the 2011-12 Run. 

\begin{figure}[htb]
\centerline{\psfig{file=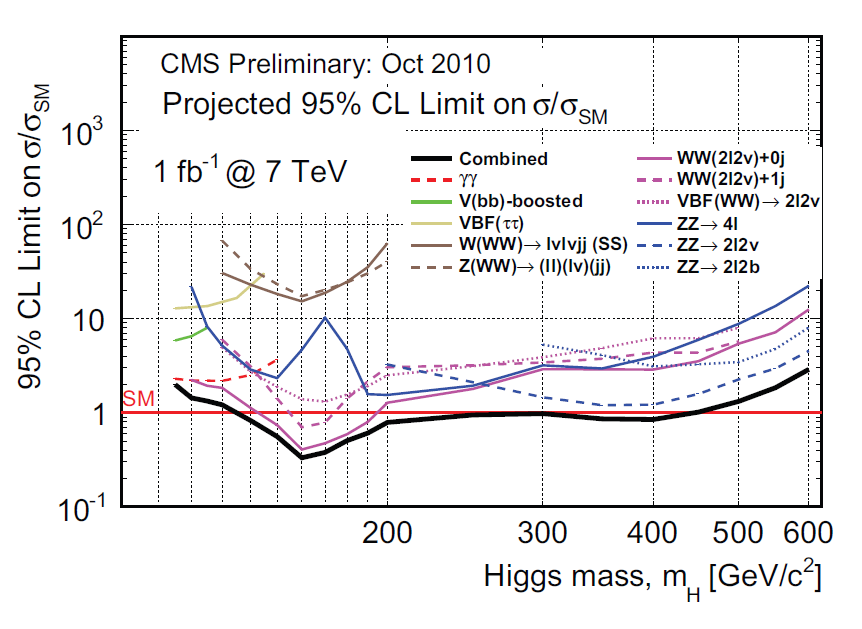, width=3.3in}}
\vspace*{8pt}
\caption{Projected exclusion limits for a SM Higgs search at $\sqrt{s}$ = 7 TeV and an integrated luminosity of 1 \ifb. Contributions of individual channels used in the overall combination are also shown. \protect\label{fig:higgs_limits}}
\end{figure} 
\begin{figure}[htb]
\centerline{\psfig{file=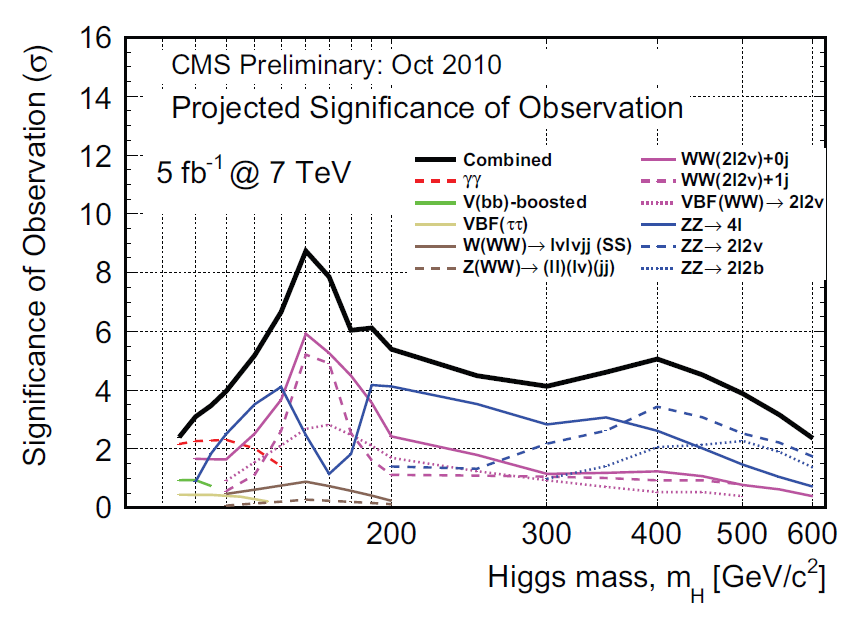, width=3.3in}}
\vspace*{8pt}
\caption{Projected expected observation significance (right) for a SM Higgs search at $\sqrt{s}$ = 7 TeV and an integrated luminosity of 5 \ifb. Contributions of individual channels used in the overall combination are also shown.\protect\label{fig:higgs_sign}}
\end{figure}

\section*{Acknowledgments}
The author wishes to thank the LHC staff for the excellent accelerator performance; his CMS colleagues for building, commissioning and continuously operating a state-of-the-art detector; and the Physics Departments at Harvard and Standford Universities for their hospitality. This article has been written with the support of the LHC Physics Center (LPC) Fellowship Program at Fermilab, US. 



\begin{thebibliography}{0}
\bibitem{bib:cms} S. Chatrchyan \etal (CMS Collaboration), JINST {\bf 3} (2008) S08004.
\bibitem{bib:l1} S. Chatrchyan \etal (CMS Collaboration), JINST {\bf 5} (2010) T03002.
\bibitem{bib:hlt1} The CMS Collaboration, ``CMS High Level Trigger'', CERN/LHCC 2007-021, LHCC-G-134 (2007), available at {\tt http://cdsweb.cern.ch/record/1043242}.
\bibitem{bib:hlt2} S. Chatrchyan \etal (CMS Collaboration), JINST {\bf 5} (2010) T03005.
\bibitem{bib:pf_lep} D. Buskulic \etal (ALEPH Collaboration), Nucl. Instr. Methods A {\bf 360} (1995) 481.
\bibitem{bib:pf} The CMS Collaboration, Physics Analysis Summary CMS-PAS-PFT-09-001, available at {\tt http://cdsweb.cern.ch/record/1194487}.
\bibitem{bib:chd_had} V. Khachatryan \etal (CMS Collaboration), ``Charged particle transverse momentum spectra
in $pp$ collisions at $\sqrt{s}$ = 0.9 and 7 TeV'', J. High Energy Phys. (2011, submitted), {\tt arXiv:1104.3547}.
\bibitem{bib:inc_jet} The CMS Collaboration, Physics Analysis Summary CMS-PAS-QCD-10-011, available at {\tt http://cdsweb.cern.ch/record/1280682}.
\bibitem{bib:inc_pho} V. Khachatryan \etal (CMS Collaboration), Phys. Rev. Lett. {\bf 106}, 082001 (2011).
\bibitem{bib:chd_B} V. Khachatryan \etal (CMS Collaboration), Phys. Rev. Lett. {\bf 106}, 112001 (2011).
\bibitem{bib:neu_B} V. Khachatryan \etal (CMS Collaboration), ``Measurement of the $B^0$ production cross section measurements in $pp$ Collisions at $\sqrt{s}$ = 7 TeV'' Phys. Rev. Lett. (2011, submitted), {\tt arXiv:1104.2892}.
\bibitem{bib:Bs} The CMS Collaboration, Physics Analysis Summary CMS-BPH-10-013, available at {\tt https://twiki.cern.ch/twiki/bin/view/CMSPublic/PhysicsResultsBPH10013}.
\bibitem{bib:onia1} V. Khachatryan \etal (CMS Collaboration), Eur. Phys. J. C {\bf 71} (2011) 1575.
\bibitem{bib:onia2} V. Khachatryan \etal (CMS Collaboration), ``Measurement of the Inclusive Upsilon production cross section in $pp$ collisions at $\sqrt{s}$ = 7 TeV'' Phys. Rev. D (2011, submitted), {\tt arXiv:1012.5545}.
\bibitem{bib:onia3} The CMS Collaboration, Physics Analysis Summary CMS-PAS-BPH-10-018, available at {\tt http://cdsweb.cern.ch/record/1345725}.
\bibitem{bib:ewk} The CMS Collaboration, Physics Analysis Summary CMS-PAS-EWK-10-005, available at {\tt http://cdsweb.cern.ch/record/1337017}.
\bibitem{bib:top} The CMS Collaboration, Physics Analysis Summary CMS-PAS-TOP-11-001, available at {\tt http://cdsweb.cern.ch/record/1336491}.
\bibitem{bib:candle} See, for example, M. Dittmar, F. Pauss, and D. Z\"{u}rcher Phys. Rev. D {\bf 56}, 7284-7290 (1997) and {\tt hep-ph/0702251}.
\bibitem{bib:lq1}  V. Khachatryan \etal (CMS Collaboration), ``Search for Pair Production of First-Generation Scalar Leptoquarks in $pp$ Collisions at $\sqrt{s}$ = 7 TeV'', Phys. Rev. Lett. (2010, accepted), {\tt arXiv:1012.4031}.
\bibitem{bib:lq2}  V. Khachatryan \etal (CMS Collaboration), ``Search for Pair Production of Second-Generation Scalar Leptoquarks in $pp$ Collisions at $\sqrt{s}$ = 7 TeV'', Phys. Rev. Lett. (2010, accepted), {\tt arXiv:1012.4033}.
\bibitem{bib:altarelli} See, for example, G. Altarelli, B. Mele and M. Ruiz-Altaba, Z. Phys. C {\bf 45} (1989) 109.
\bibitem{bib:wprime_e} V. Khachatryan \etal (CMS Collaboration),  Phys. Lett. B {\bf 698} (2011) 21-39.
\bibitem{bib:wprime_mu} S. Chatrchyan \etal (CMS Collaboration), ``Search for a $W^\prime$ boson decaying to a muon and a neutrino in $pp$ collisions at $\sqrt{s}$ = 7 TeV'', Phys. Lett. B (2011, submitted), {\tt arXiv:1103.0030}.
\bibitem{bib:zprime} S. Chatrchyan \etal (CMS Collaboration), ``Search for Resonances in the Dilepton Mass Distribution in $pp$ collisions at $\sqrt{s}$ = 7 TeV'', J. High Energy Phys. (2011, accepted), {\tt arXiv:1103.0981}.
\bibitem{bib:stopped_gluino}  V. Khachatryan \etal (CMS Collaboration), Phys. Rev. Lett. {\bf 106}, 011801 (2011).
\bibitem{bib:rhic} B. Alver \etal (PHOBOS Collaboration), Phys. Rev. C {\bf 81} (2010), 024904.
\bibitem{bib:ridge}  V. Khachatryan \etal (CMS Collaboration),  J. High Energy Phys. {\bf 09} (2010) 091.
\bibitem{bib:higgs} The CMS Collaboration, ``Projected sensitivity for SM Higgs boson searches'', available at {\tt https://twiki.cern.ch/twiki/bin/view/CMSPublic/PhysicsResultsHIG}.
\end{thebibliography}
\end{document}